\documentclass[journal]{IEEEtran}
 \pdfoutput=1

\usepackage{balance}
\usepackage{graphicx}
\usepackage [noadjust]{cite} 
\usepackage{bm}  
\usepackage{amsmath}
\usepackage{amssymb}
\usepackage{enumerate}
\usepackage{stfloats}
\usepackage{cases}
\usepackage{epstopdf}
\usepackage{soul,color} 
\usepackage{hyperref}
\usepackage[percent]{overpic}
\usepackage{tabularx}
\usepackage[table]{xcolor}
\usepackage{multirow}
\usepackage{booktabs}  
\usepackage{tabu} 

\usepackage{algorithm}
\usepackage{algpseudocode}%

\newcolumntype{L}[1]{>{\hsize=#1\hsize\raggedright\arraybackslash}X}%
\newcolumntype{R}[1]{>{\hsize=#1\hsize\raggedleft\arraybackslash}X}%
\newcolumntype{C}[1]{>{\hsize=#1\hsize\centering\arraybackslash}X}%

\newcommand{\hbold}{\boldsymbol{h}}

\newcommand{\wbold}{\boldsymbol{w}}

\newcommand{\abold}{\boldsymbol{a}}

\newcommand{\rbold}{\boldsymbol{r}}

\newtheorem{property}{Property}

\begin{document}
	\raggedbottom
	\allowdisplaybreaks

     \title{Towards Near-Field 3D Spot Beamfocusing: Possibilities, Challenges, and Use-cases}
    
	\author{Mehdi Monemi, \textit{Member}, IEEE, Mohammad Amir Fallah, Mehdi Rasti, \textit{Senior Member},
IEEE, Matti Latva-Aho, \textit{Fellow}, IEEE,  and Merouane Debbah, \textit{Fellow}, IEEE
		\thanks{\ }
	}
    	

	\maketitle
\begin{abstract}
Spot beamfocusing (SBF) is the process of focusing the signal power in a small spot-like region in the 3D space, which can be either hard-tuned (HT) using traditional tools like lenses and mirrors or electronically reconfigured (ER)  using modern large-scale intelligent surface phased arrays. ER-SBF (simply called SBF) can be a key enabling technology (KET) for the next-generation 6G wireless networks offering benefits to many future wireless application areas such as wireless communication and security, mid-range {\color{black}high-power and safe} wireless battery charging systems,  medical and health, physics, etc.
Although near-field HT-SBF and ER-beamfocusing have been studied in the literature and applied in the industry, there is no comprehensive study of different aspects of SBF and its future applications, especially for nonoptical (mmWave, sub-THz, and THz) electromagnetic waves in the next generation wireless technology, which is the aim of this paper. The theoretical concepts behind SBF, different antenna technologies for implementing SBF, employing machine learning (ML)-based schemes for enabling channel-state-information (CSI)-independent SBF, and different practical application areas that can benefit from SBF will be explored.

\end{abstract}
	
	
\thispagestyle{empty}

\section{Introduction}
\label{sec:introduction} 
The terms beamforming and beamfocusing have different interpretations in the literature. Therefore, first, we clarify the definitions of these terms as they are used in this article.
\textbf{Beamforming} is a technique that directs a wireless signal towards a specific receiving device, rather than spreading the signal omnidirectionally. 
\textbf{Beamfocusing} is a special kind of 3D-beamforming where most of the radiated power is concentrated in a confined focal region around a point in the 3D space, defined by both angular (azimuth and elevation angles) and radial domains. This is unlike traditional 2D beamforming, which only considers the angular domain. We use the term spot beamfocusing (\textbf{SBF}) for the special case where the focal region of beamfocusing is very small, i.e., the power is focused in a spot-like region around the focal point \cite{monemi20236GFresnel}.
In this paper, we study various issues relating to SBF for electromagnetic (EM) signals in the near-field region.
SBF has many potential applications not only in wireless communication and wireless power transfer (WPT) but also in health and medical sensing (e.g., stimulating specific neurons through neuromodulation \cite{krames2018neuromodulation}), 
semiconductor and THz technology (e.g., high speed turning on/off nano switch arrays through casting spot-like power on each switch element \cite{ou2020tunable}), 
etc.  
Implementing the SBF in these various applications is generally realized as hard-tuned (HT), or electronically reconfigured (ER). In hard-tuned SBF (\textbf{HT-SBF}),  there exists no soft control on the location of the focal point, while electronically reconfigurable SBF (\textbf{ER-SBF}) schemes can adjust the focal point by softly configuring the radiated beam of the aperture. In general, due to the limitations of HT-SBF, ER-SBF schemes are preferred which are elaborated in what follows.

HT-SBF is generally implemented through lenses, mirrors, or gratings.
These technologies face challenges such as diffraction, and aberrations, alignment mismatch \cite{bayanna2015membrane}, as well as fixed focal points. There exist limited mechanical methods to change the desired focal point (DFP) location, such as employing stepper motors. However, they are rather slow and non-flexible, have limited control over the DFP position, and require regular maintenance services. 
A more flexible and potentially cheaper option is to implement ER-SBF through intelligent apertures.
ER-SBF requires extremely large-scale apertures \cite{goodman2017introduction} and can be realized through different antenna technologies such as {\color{black}conventional phased array antennas (CPAs)} 
, dynamic metasurface antennas (DMAs) 
and holographic MIMO (HMIMO) surfaces. 
While HT-SBF and ER-beamfocusing have been widely studied in the literature so far and implemented in many optical and non-optical EM applications, there exists no comprehensive study of different aspects of smart ER-SBF systems and their future applications, especially for nonoptical (mmWave, sub-THz, and THz) EM waves in the next generation wireless technology. Hereafter, we simply use SBF for the term ER-SBF.
In this paper, we are going to address the following key questions:
\begin{itemize}
    \item  How can near-field ER-beamfocusing be extended to near-field SBF? Which beamforming structures are efficient for realizing SBF?
    \item
    What are the challenges, advantages, and disadvantages associated with deploying SBF using various smart antenna technologies? Additionally, how do various design parameters 
    impact the performance of SBF?
    \item How can machine learning techniques be leveraged to enhance the performance of SBF systems?
    \item What are potential SBF applications in various domains?
\end{itemize}

\section{Near-field (Fresnel) SBF}
\label{nearfield_ER_SBF}
\subsection{Beamfocusing in different propagation zones}
\label{sec:farnear}

\begin{figure*}
		\centering
		\includegraphics [width=508pt]{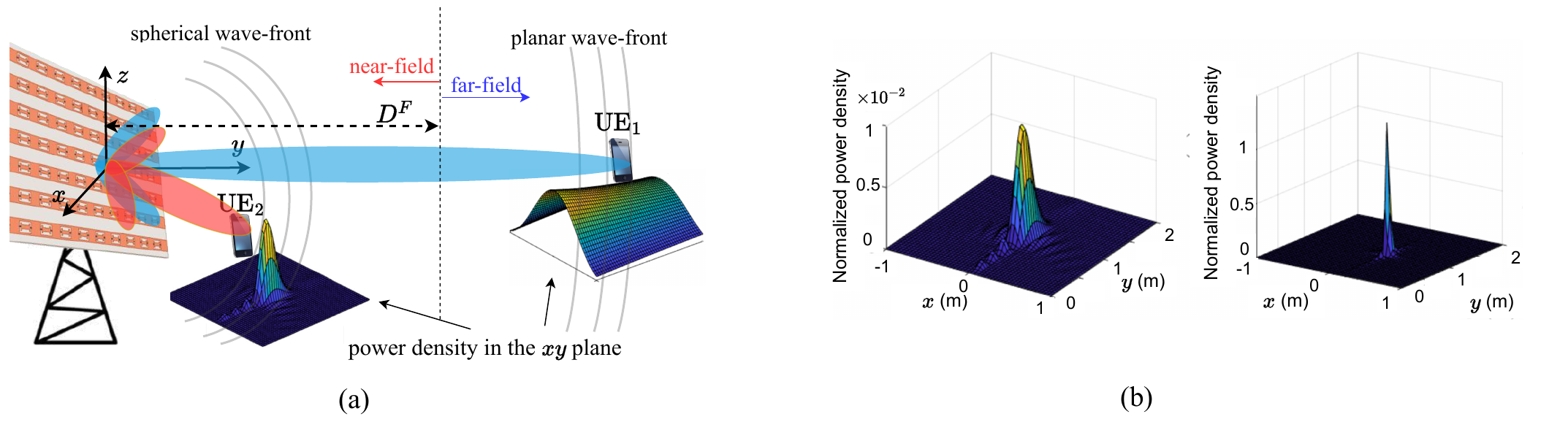} \\
		\caption{(a) 2D  beamforming is feasible for far-field in angular domain for UE1 with no radial domain beamfocusing in the $y$-axis, and 3D  beamfocusing is feasible for UE2 in the near-field. (b): Near-field spacial power density on the $xy$ plane for an aperture located on $xz$ plane and a DFP at (0,1,0). The antenna elements are half wavelength apart and the frequency is 28GHz. The left and right figures correspond to $6 \times 6$ SPA and $60 \times 60$ ELAA respectively.}
		\label{fig:far-field}
\end{figure*}

To begin with, we explore the case of signal propagation and beamforming in the far-field region. This is defined as the region where the distance from the measuring point to the transmitting aperture exceeds the Fraunhofer limit $D^F$. The Fraunhofer limit can be approximated by $D^F\approx 2D^2/\lambda$ on the boresight of the antenna \cite{selvan2017fraunhofer} where $D$ is the diameter of the aperture and $\lambda$  is the wavelength. 
 The realization of sharp and even pencil beamforming is achievable through far-field beamshaping schemes in the \textit{angular domain}, but the implementation of \textit{3D beamfocusing} requires the \textit{radial domain} beamforming as well, which is not feasible through a \textit{single} transmitting aperture due to the monotonically decreasing power level in the radial domain in the far-field region.
 This is shown for the second UE of Fig. \ref{fig:far-field}-a. 
Unlike the far-field scenario, where 3D beamfocusing is impossible with a single aperture, in the near-field we may focus the radiated signal power around a desired point through a single phased array antenna. Beamfocusing is feasible in \textit{radiative near-field} region wherein the distance from the aperture ($r$) lies in the region $D^N<r<D^F$ where $D^N$ is a small distance very close to the antenna (usually lower than a wavelength) beyond which the radiative active power dominates the non-radiative reactive power. 
In the far-field region, 
a change in $r$ leads to an equal change in the arrival phase of signals from all array elements, resulting in no directivity change in this direction. However, this does not hold in the near-field scenario due to the spherical wave-front.
More specifically, by applying appropriate phase shifts to each antenna element of the phased array, the received signals radiated from each antenna element can be constructively (i.e., coherently) added at the DFP; however, since different channel phase shifts are experienced for each antenna element at points around the DFP, power decay might be experienced at these points, leading to 3D beamfocusing at the DFP. This is illustrated for UE2 in Fig. \ref{fig:far-field}-a. 

\subsection{From beamfocusing to SBF}
Consider a single phased-array aperture. 
For a given beamforming vector $\wbold$, let define the beamfocusing radius (\textbf{BFR}) denoted by $R$  at some reference plane $S$, as the radius of the circle $S_R$ located on the reference plane and centered at DFP which contains a fraction $\eta$ of the total radiating power in the reference plane $S$ \cite{monemi20236GFresnel}.
    For example, a BFR value corresponding to $\eta=0.9$ implies that a circle of radius BFR centered at DFP contains 90\% of the total radiating power at the reference plane.
    Considering a sphere of radius $R$ at the DFP, it can be shown that when the near-field impact becomes dominant,  the power level fades outside the sphere in all directions, even when getting close to the aperture \cite{monemi20236GFresnel}. 
    This implies that the focal region in the near-field can be imagined as a sphere centered at the DFP.
    We are specifically interested in obtaining an electronically controlled spot-like beamfocusing (i.e., ER-SBF) with a high power density and minimal BFR (close to zero) at the DFP. This is the key point in utilizing this technology in many practical applications, overtaking the benefits of many traditional beam spotting technologies such as lenses, mirrors, and gratings.

To achieve near-field SBF, a high value of the ratio $D/\lambda$ is required \cite{goodman2017introduction}, which can be accomplished through decreasing $\lambda$ as well as increasing $D$. The former is achieved through transitioning to mm-wave/THz frequencies. The latter can be realized by using extremely large-scale antenna-arrays (ELAAs) rather than small-scale phased-arrays (SPAs).
For instance, the spatial power distribution in the $xy$ plane around the DFP at (0,1,0) for an aperture located at $xz$ plane and centered at the origin is illustrated in Fig. \ref{fig:far-field}-b for two scenarios: a $6\times 6$ SPA and a $60\times 60$ ELAA operating at 28 GHz corresponding to the left and right diagrams. The figure demonstrates that the SPA enables beamfocusing, however, the SBF can only be achieved by the ELAA, which attains a much higher power intensity and a much lower BFR at the focal point, as well as much smaller side lobes elsewhere. The SBF problem for a phased-array antenna requires finding the beamforming coefficients corresponding to minimum BFR at the DFP; this problem is generally non-convex, NP-hard, and intractable; however for ELAAs in the Fresnel region, which is the case for realizing SBF, the problem is equivalent to finding the solution to the problem of maximizing the received power at the DFP, which is tractable and less complex than the original problem (\cite{monemi20236GFresnel}). It should be noted that the realization of a sharp focal point requires that $r \ll D^F$ \cite{smith2017analysis}. This means that SBF is not feasible at regions close to the boundary of far-field and near-field.  


\subsection{Application areas of SBF through ELAAs}
\label{sec:applications}
There exists a variety of applications in physics, wireless communication, WPT, medical and health, etc., which can be highly benefited from SBF. 
In what follows we briefly introduce some of these applications. 

\textbf{\textit{Ultra-high-speed (UHS) wireless communication}}:
 SBF can provide the UE with UHS data transfer, much faster than achieved through conventional beamforming/beamfocusing schemes. 
 The reason behind that is manifold. From the UE side, casting of all transmitted power to the exact UE location point {\color{black} with minimum power leakage to the non-UE located points in the surrounding environment results in much higher SNR than conventional beamforming schemes} leading to ultra-high Shannon's channel capacity. From the network side, {\color{black} almost the whole frequency spectrum (with minimal interference) can be explicitly reused by each and every UE located in different spots in the network, leading to full-scale 3D spatial spectrum multiple access (SDMA). This is due to the asymptotic orthogonality in the near-field region expressed in the next section as {\bf {Property 1}}}. 

  \textbf{\textit{Mid-range safe {\color{black}and high-power} WPT}}:
  Mid-range WPT for wireless battery charging and energy transfer through SBF  will potentially be a KET in the next generation of wireless networks. Nowadays, the industry offers low-range WPT battery charging facilities that require the UE to be in the non-radiative near-field region, very close to the power transmitting antenna. {\bf Health hazards} are the main barrier to extending the wireless battery charging distance from low-range distances (in the order of several centimeters) to mid-range distances (in the order of several meters).
  {\color{black} In the conventional near-field electric-field coupling or magnetic inductive WPT schemes, the inductive/coupling efficiency highly decreases as the distance increases, making these technologies mostly efficient for short-range WPT applications.
  On the other hand, far-field techniques are exposed  to harmful radiation outside the
intended target zones, leading to health hazard for high-power WPT applications if some body tissues are placed between
the radiating antenna and the energy absorbing device. 
Near-field SBF addresses these shortcomings by offering a highly focused and efficient solution. In essence, near-field spot beamfocusing offers a unique combination of safety, efficiency, and spatial selectivity for high-power mid-range WPT applications by minimizing the power leakage in non-target zones in 3D space. Further research is required to consider the system complexity and investigate the sensitivity of the SBF WPT performance measure in various propagating environments, as well as to keep an eye on the electromagnetic field (EMF) exposure not exceeding standard limits in the surrounding environment.
}

    {\color{black}  
     \textbf{\textit{Leveraging physical layer security for data transfer}}:
 SBF can greatly improve the physical layer security level of data transfer. In particular, dense IoT devices with limited resources might be easily exposed to eavesdropping by unauthorized receivers. By focusing all transmitted data energy exclusively on the intended receiver, SBF effectively places all eavesdroppers within the blind zone. Consequently, SBF thwarts eavesdroppers’ ability to intercept the data at the physical layer, even if they are situated between the transmitter and the intended receiver. To numerically illustrate
this, consider the simulation scenario similar to that relating to Fig. \ref{fig:plot1}-a a $N$-element uniform planar array (UPA) operating at frequency 28GHz with equipped with two fully connected RF chains dedicated for beamfocusing at two DFPs is considered. The interelement spacing (the distance between neighboring antenna elements) is considered half a wavelength. The noise power is normalized to 1 W and the transmit power is adjusted such that the SNR is $\gamma=10$ dB at each of the DFPs. Assuming a minimum required SNR for the successful decoding probability of $\gamma^{\mathrm{th}}=5$dB (corresponding to the region inside the red closed curves), the physical layer security corresponding to $\gamma \leq \gamma^{\mathrm{th}}$ is depicted in Fig. \ref{fig:physical_security} for $5\times 5$, $15\times 15$ and $60\times 60$ UPAs. It is seen how going toward SBF with ELAAs enlarges the physical layer secure region, prohibiting eavesdroppers from decoding the signal at the physical layer.
\begin{figure}
    \centering
    \includegraphics[width=254pt]{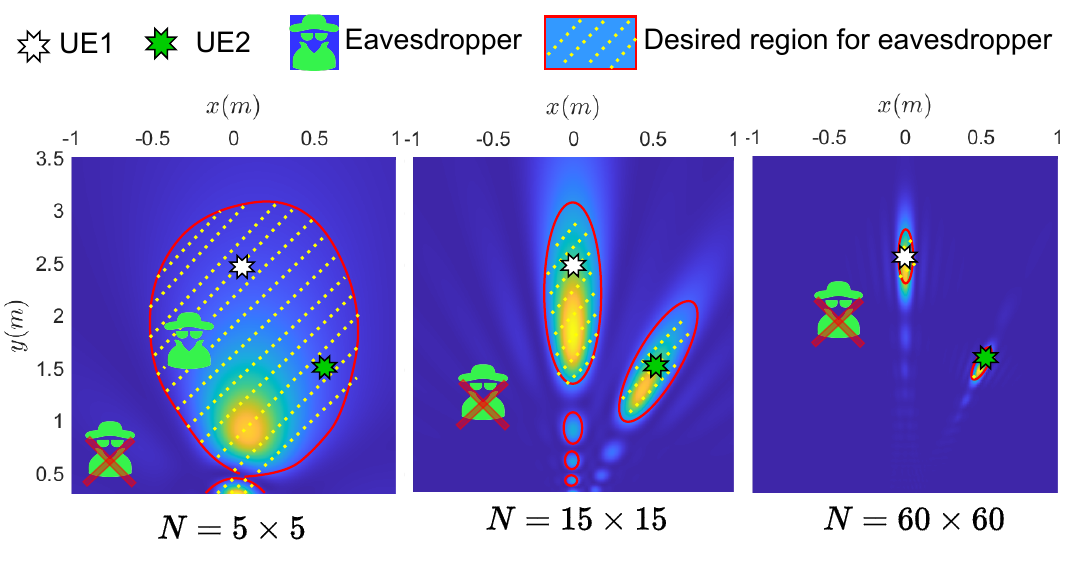}  \caption{{\color{black}Illustration of the physical layer security region for a $N$-element MIMO communication using UPA located on $xz$ plane centered at the origin.}}
    \label{fig:physical_security}
\end{figure}
 }

     \begin{figure}
    \centering
    \includegraphics[width=254pt]{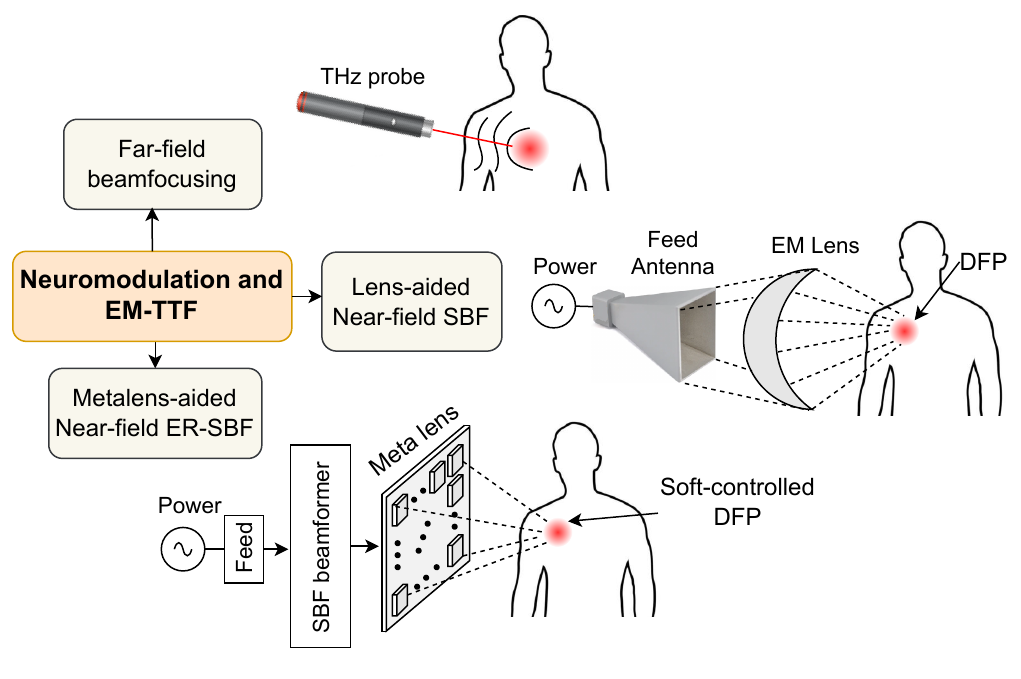}
    \caption{{\color{black}Three schemes for implementing EM-based neuromodulation and tissue thermal focusing.}}
    \label{fig:neuro}
\end{figure}

 {\color{black}
 \textbf{\textit{EM neuromodulation and tissue thermal focusing}}:
\textbf{EM neuromodulation} is a technique that modulates neural activity through the targeted delivery of EM stimuli to specific neurological sites within the body\cite{krames2018neuromodulation}. \textbf{EM Tissue Thermal Focusing (EM-TTF)} is a method of concentrating EM energy to induce a desired thermal profile in a target tissue. Both EM-neuromodulation and EM-TTF can be achieved using either 2D far-field beamfocusing at sub-THZ/THz frequencies\cite{liu2023recent}, or 3D near-field beamfocusing at sub-20GHz\cite{abd2024design,mahmoud2022design}.
In the far-field approach, a laser-like EM beam is transmitted through a THz probe towards the target tissue. The frequency is selected to ensure penetration through intervening tissues and subsequent absorption by the target. However, far-field neuromodulation is subject to limitations that can be addressed through near-field beamfocusing. Specifically, energy dissipation in non-target tissues and potential absorption by tissues with similar absorption properties as the target can compromise the effectiveness of the far-field technique.
Recent research has explored the application of 3D near-field beamfocusing for in-body applications.
For example, in \cite{abd2024design}, the beam radiated by a feed antenna at $f=2.45$GHz is passed through a dielectric lens to effectively be focused inside the human body. 
The authors of \cite{mahmoud2022design} presented a multi-resonance flexible antenna array applicator  at $f=4.5$GHz designed for breast cancer hyperthermia treatment. This array is optimized with a hybrid algorithm to focus heat on tumors.
To further enhance control over the focused EM field, static lenses can be replaced with meta-lenses, as illustrated in Fig. \ref{fig:neuro}. It is essential to note that the optimal frequency and antenna structure for near-field beamfocusing using meta-lenses require careful investigation and are subject to certain constraints. While lower frequencies generally lead to increased skin depth, they also result in larger antenna dimensions. For example, a $50\times 50$ ELAA meta-lens employed at $f=15$GHz with interelement spacing of half a wavelength is about $50\times 50\mathrm{cm}^2$. Additionally, scattering, reflection, and polarization losses within biological tissues significantly impact design parameters and necessitate thorough analysis.
}

     \textbf{\textit{THz switching}}:
     {\color{black}
     THz switching refers to the capability of selectively activating or deactivating a specific switch within an array on a picosecond timescale. This can be achieved by directing THz EM radiation toward the desired switch. In the case of single-layer 2D switches, far-field pencil-like beams can be used to precisely steer and focus the THz stimulus \cite{ou2020tunable}. For multi-layer switches, 3D spatial SBF can be employed. Steering the spot-like energy focus in the angular domain and radial domain can enable the activation/deactivation of switches within the same or different layers respectively.
     }


   
\section{Technical Issues and Practical Considerations for Achieving SBF}    
\subsection{SBF from a beamforming implementation perspective}
Noting that SBF necessitates an extensive array of elements, the implementation of beamforming can result in intricate software and hardware structures.
Consequently, the proposed design should prioritize simplicity. The pivotal question that emerges is: What type of beamformer is best suited for implementing SBF? The answer to this question lies behind the following property:
\begin{property}
    The normalized array responses for any given two points $\rbold_1 \neq \rbold_2$ denoted by $\abold(\rbold_1)$ and $\abold(\rbold_2)$ have asymptotic
orthogonality in the near-field  if the
number of antennas $N$ is sufficiently large. This property is mathematically expressed as follows\cite{Liu2023near}:
\begin{align}
    \label{eq:asymp}
    \lim_{N\rightarrow \infty} \frac{1}{N} |\abold^{\mathrm{T}} (\rbold_1) \abold^*(\rbold_2)|=0
\end{align}
\end{property}
To put it concisely, Property 1 states that for ELAAs having a very large $N$, if we consider any beamforming vector such as the simple maximum ratio transmission (MRT) precoding corresponding to point $\rbold_1$,   this has negligible interference on any other point $\rbold_2\neq \rbold_1$. This leads to specific implications for beamformer design in the following two cases:

{\bf $\circ$ Single focal SBF}: In this case, according to Property 1, one low-complexity option is to consider the simplest MRT beamforming scheme in the analog domain through a single RF chain pushing all signals to be added coherently at the DFP. This means that there is no need to be concerned about the beamwidth, side-lobes, and other undesired effects through more intricate beamforming schemes, and these unwanted results asymptotically disappear as $N$ becomes very large. This eliminates the computationally expensive procedures required for complicated beamforming schemes considering the extremely large cardinality of the channel matrix.

{\bf $\circ$ Multi focal SBF}: Consider a beamformer aiming to create $M$ focal points, corresponding to data streams for $M$ users at different location points. According to Property 1, when the number of array elements is sufficiently large, the interference signal from spatially separated focal points becomes negligible. This obviates the need for complex digital domain algorithms such as {\color{black}minimum mean square error} (MMSE) or {\color{black}zero-forcing} (ZF), which would otherwise require implementation across thousands of array elements. Assuming a sufficiently large $N$, a low-complexity beamformer can be designed. This structure employs  $M$ RF chains followed by a fully connected MRT analog beamformer. Despite its minimal complexity, this approach exhibits negligible performance degradation compared to other high-complexity optimal fully digital or hybrid beamformers. 

\subsection{The effect of interelement spacing and array size}
\begin{figure}
\centering
        \begin{tabular}{c}

        \includegraphics [height=135pt]{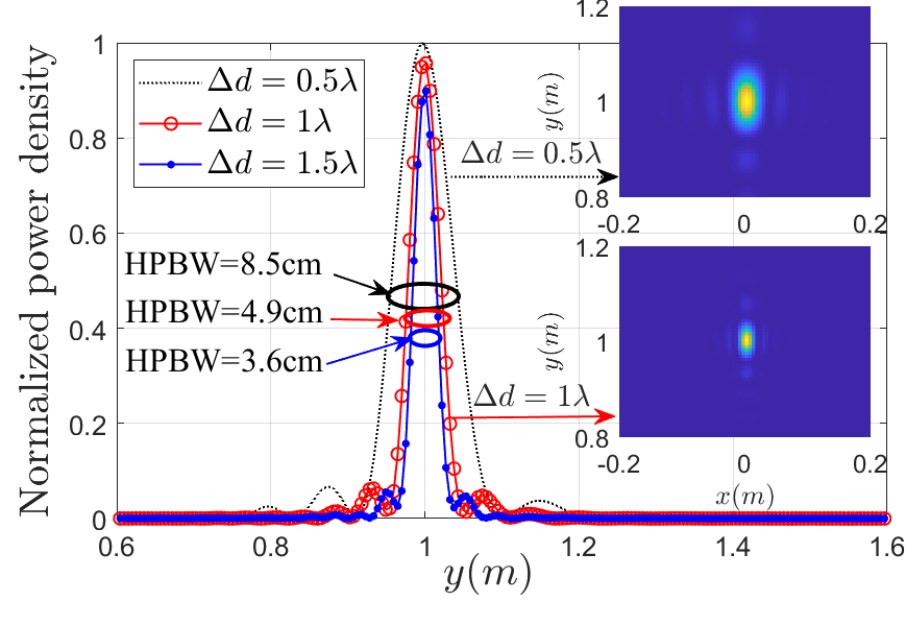} 
        \vspace{-5pt}
        
        \\
         {\small (a)}
        \\
        
        \ \ \ \ \ \includegraphics [height=115pt]{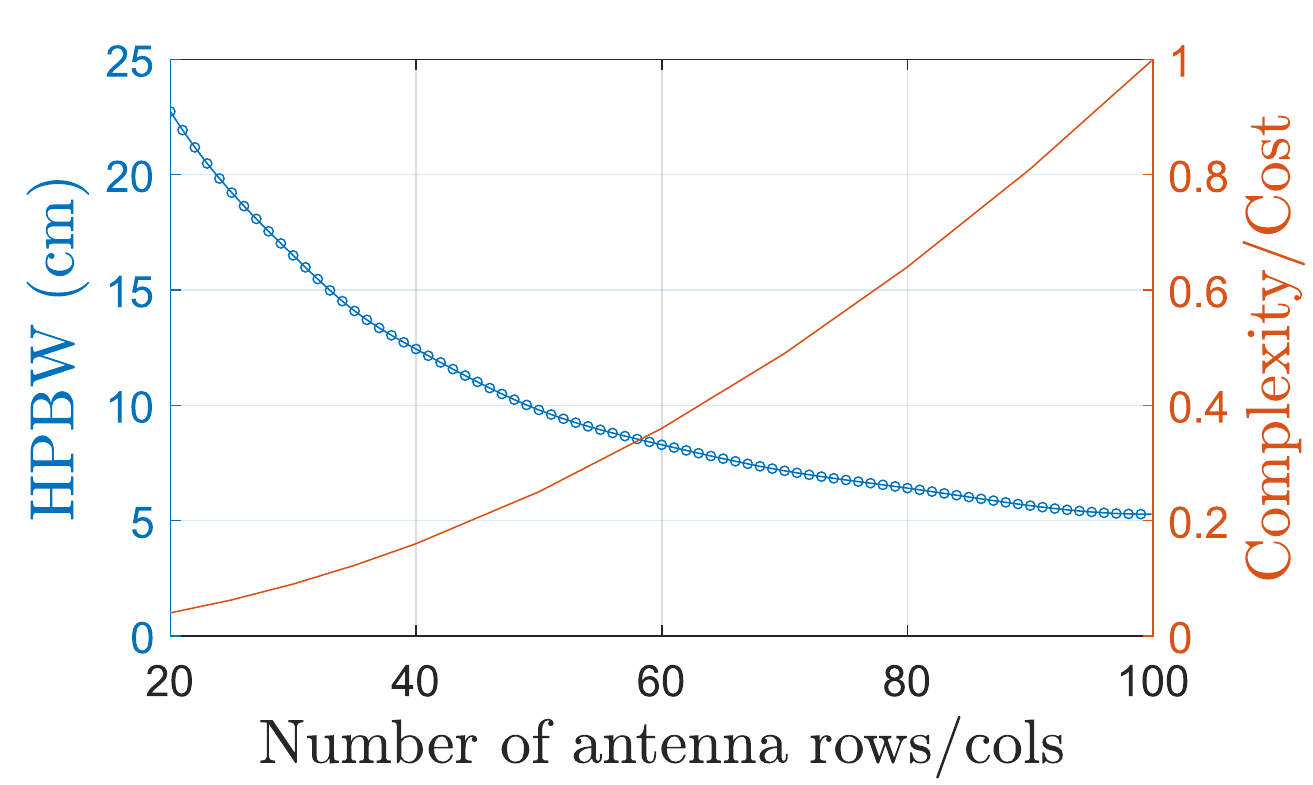} 
          \vspace{-5pt}
        
        \\
        {\small (b)}
        \\
        
        \end{tabular}
     
		\caption{{\color{black}Realization of SBF through extremely large-scale UPA. (a): Power density versus distance ($y$) for $\Delta d/ \lambda \in\{0.5,1,1.5\}$ using a $N=60 \times 60$ ELAA. (b): Complexity/cost, as well as HPBW versus the number of antenna rows/cols (i.e., $\sqrt{N}$) of the UPA for $\Delta d=0.5 \lambda$.}}
\label{fig:plot1}
\end{figure}

Various aspects of the implementation of SBF can be explored. For instance, {\color{black}let $\Delta d$ represent the distance between neighboring elements of the ELAA which is referred to as interelement spacing.}  If we consider a given number of array elements for the ELAA, a key question is how to determine the optimal value of $\Delta d/ \lambda$ for achieving the best SBF. In practice, this ratio is usually selected as 0.5 for non-holographic CPA/DMA apertures in applications involving far-field/near-field beamforming.
However, the SBF performance is sensitive to the ratio $\Delta d/ \lambda$, which affects it differently than regular beamforming applications.
Considering $\Delta d/ \lambda \in \{0.5,1,1.5\}$ as an example, Fig. \ref{fig:plot1}-a illustrates the normalized absolute value of the power density in the $y$ direction for {\color{black}an extremely large-scale UPA consisting of $60 \times 60$ elements} located on the $xz$ plane centered at the origin, operating at frequency 28GHz, and whose DFP is at (0,1,-0.5) which is 1.12m away from the center of the aperture. 
It is seen that the maximum peak power received at the DFP occurs when $\Delta d/ \lambda=0.5$. This is because increasing $\Delta d/ \lambda$ makes some array elements farther from the DFP, which reduces channel gains and the aggregate peak power. However, higher values of $\Delta d/ \lambda$ result in lower BFR. Here, instead of the BFR, we have specified the more simplistic case of half-power beam width (HPBW). For example, for a fixed value of $\lambda$, increasing $\Delta d/ \lambda$ increases the aperture diameter, which enables higher near-field benefits of achieving sharper beam focus and lower BFR. Therefore, a tradeoff between high received power and low BFR should be considered.  For instance, changing $\Delta d/ \lambda$ from 0.5 to 1 decreases the measured power density at DFP by only 4\%, while the HPBW is highly improved from 8.5cm to 4.9cm. A higher aperture diameter can also be achieved by using a higher number of array elements instead of increasing $\Delta d/ \lambda$. However, this requires more expensive hardware and complex beam-tuning software, which might not always be preferred. It should be noted that increasing $\Delta d/\lambda$ from 0.5 has the benefit of creating a sharply focused beam with fewer antenna elements, however, this makes the antenna sparse, which in turn, may lead to the occurrence of additional focal points due to near-field grating lobes. Another technical aspect of SBF is the elaboration of cost per how small the SBF can be realized. Fig. \ref{fig:plot1}-b depicts the performance of SBF in terms of HPBW versus the number of antenna rows/cols for the simulation scenario same as that in Fig. \ref{fig:plot1}-a, but with $\Delta d=0.5 \lambda$, and considering a various number of antenna elements. Assuming that the complexity and cost of the SBF structure are proportional to the number of antenna elements, the figure shows how increasing the cost leads to a smaller focal region.  

\subsection{How different antenna technologies impact SBF}
SBF can be realized through CPAs, DMAs, or HMIMO surfaces. In practice, each technology leads to different characteristics of the realized SBF.
{\color{black}CPAs use discrete and separate components to control the phase and amplitude of each antenna element, resulting in bulky and complex systems}; CPAs are typically suitable for SBF through large apertures at RF frequencies. At RF bands, large CPAs employing high-power transmitters can generate a focal point at distances ranging from tens to hundreds of meters, potentially with higher focused power levels at the DFP compared to other technologies {\color{black}(due to the higher effective antenna surface)}. DMAs on the other hand can be employed to produce highly low-cost, mass-producible reconﬁgurable apertures that can be manufactured in dense and large-scale ultra-thin microstrip structure \cite{kumar2022comprehensive}. DMAs employ arrays of programmable meta-atoms that can be manufactured in various sizes, enabling SBF for a variety of applications in mmWave, sub-THz, and THz frequency bands. To achieve perfect SBF at the precise location of DFP with minimal BFR, each element of the metamaterial ELAA should be provided with analog full-scale phase control or discretized control with low quantization error. In this regard, although binary DMAs are applied for communication, they can not be employed for creating 3D spot-like beams.

Finally, HMIMO surfaces offer 3D beamshaping using holographic surfaces and EM signal technology \cite{gong2023holographic}. They comprise densely packed elements (typically with sub-wavelength spacing) and can be metamaterial-based or conventional. HMIMO antennas achieve beamshaping via various holographic patterns on their apertures. By applying a recorded hologram pattern and specific EM signals via one or multiple feeds (with fewer feeds than radiating elements), they transmit a beam with the desired directivity pattern. Designing the hologram pattern carefully enables the formation of SBF at a specific DFP. HMIMO surfaces employ unique structures for amplitude/phase tuning, such as leaky-wave antennas (LWAs) \cite{gong2023holographic}. This leads to the replacement of expensive and power-hungry RF devices, resulting in more cost-effective solutions. While HMIMO surfaces achieve precise beamfocusing at the DFP location, the limited number of feeds and restricted directivity control mechanisms result in less flexible soft-tuning DFPs compared to other technologies.

In summary, CPAs are preferred when the DFP has high dynamics, or high-power SBF at longer distances is required. Conversely, if high static resolution, as well as low cost and compactness of the antenna, are taken into account, holographic surfaces are preferred. 
DMAs, on the other hand, combine the benefits of both CPAs and HMIMO surfaces.

{\color{black}
Similar to active antenna arrays, extremely large-scale (EL) reconfigurable intelligent surfaces (RISs) can also be employed to direct the beam radiated from a transmitter into a spot-like DFP in the near-field region of the EL-RIS. However, the performance might be lower due to the power leakage from the transmitter to the RIS as well as higher path loss emanating from longer signal propagation from the transmitter to RIS and RIS to the DFP. Reducing the transmitter(feed)-RIS distance and using high-directivity feeds like horn antennas can reduce path loss and power dissipation respectively.
}
\subsection{CSI estimation for SBF}
SBF requires the exact CSI of all array elements of ELAA. This can be achieved in one of two ways: (a) similar to the assumption made in HT-SBF systems, if the exact location of the DFP, as well as the exact channel model, are known, exact CSI is obtained according to the model, or (b) by employing near-field CSI estimation techniques.
Existing channel estimation methods for ELAA and massive MIMO antennas rely heavily on channel sparsity in the angular domain, which is only valid for planar wavefront assumptions in far-field propagation. Recently, the authors of \cite{cui2022channel} proposed a polar-domain sparse representation of the channels with a compressive ratio of about 50\% for a 256-element antenna. However, this idea only works for 1D linear arrays and does not apply to the 2D ELAAs employed for SBF.

In the near-field, the channel coefficients matrix is not sparse due to the spherical wavefront, even for the simple case where no multi-path propagation exists. 
Considering this, together with the extremely large number of antenna elements required for SBF, the conventional channel estimation methods employed for massive MIMO are impractical to near-field ELAAs due to very high pilot overhead and processing overload.
Two ML-based schemes might be applied to overcome this problem: (a) applying ML to estimate CSI \cite{zhang2023near}, or (b) applying CSI-independent ML algorithms to adaptively focus the beam at the DFP \cite{monemi20236GFresnel}. This is further elaborated in the following.



\subsection{Application of ML and Transfer Learning for SBF}
\label{sec:ML}
\begin{figure}
		\centering
		\includegraphics [width=254pt]{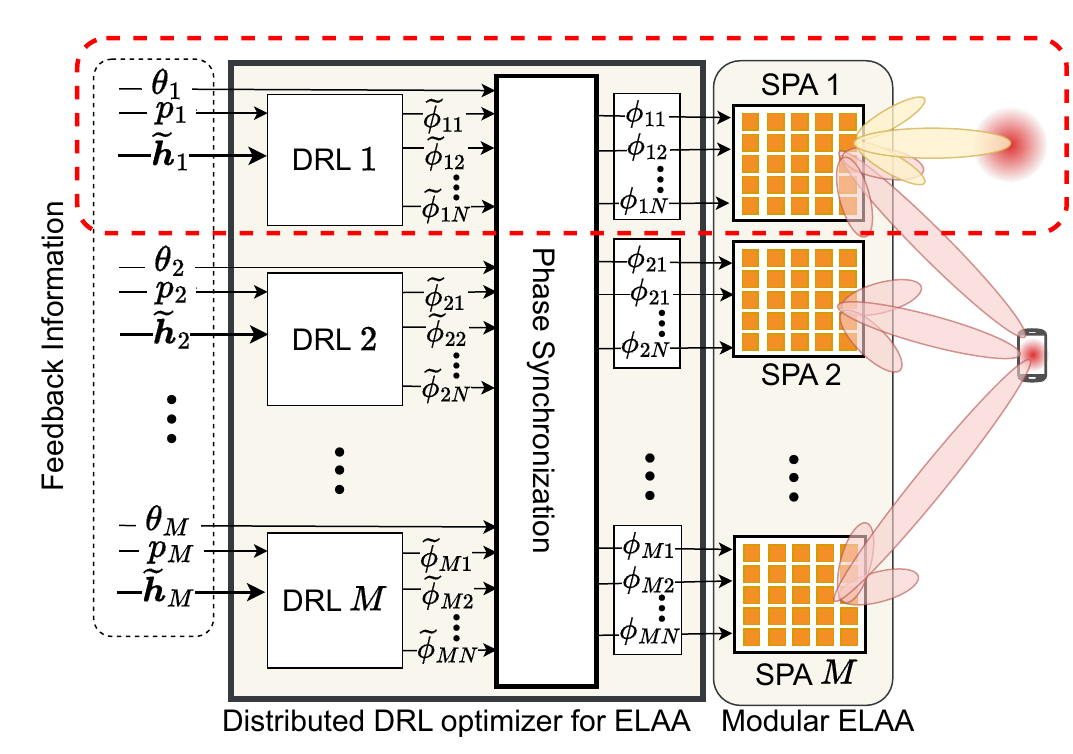} \\
  \caption{Implementation of ML-based SBF through an ELAA consisting of a set of SPAs.}
		\label{fig:ML}
\end{figure}
ML can enhance the performance of SBF systems, enabling adaptive and precise SBF for cases where the CSI is unknown or inexact. 
Consider the dashed red block in Fig. \ref{fig:ML} and let $\widetilde{\phi}_1=\phi_1$. This block can serve as an ML structure for near-field beamfocusing through a deep reinforcement learning (DRL) optimizer for an $N$-element SPA. The DRL block receives the CSI estimation (if available), as well as the power level measured by the UE at each learning epoch, and then it calculates the next step beamforming vector $\boldsymbol{\phi}=[\phi_1,\phi_2,..., \phi_N]$. The procedure continues until the output response converges and the DRL learns the optimal vector $\boldsymbol{\phi}^*=[\phi_1^*,\phi_2^*,..., \phi_N^*]$ that corresponds to the maximum focused power.
The proposed structure only works for beamfocusing using SPAs. However, SBF requires ELAAs with extremely large action space, where conventional ML schemes cannot handle the computational complexity in the original form. For example, a $60\times 60$ ELAA with 4-bit phase shifters has an action space with a cardinality of $16^{3600}$, which is too large and unaffordable for a single DRL.
A practical scheme is to split the ELAA into a set of SPAs each equipped with a DRL optimizer to collaboratively (i.e., synchronously) focus the beams and form the overall spot-like highly concentrated power at the DFP \cite{monemi20236GFresnel}. For example, a $3600$-element ELAA can be split into 100 SPAs each having $6\times 6$ antenna elements. The schematic of the proposed structure is depicted in Fig. \ref{fig:ML}, wherein a $M\times N$ element ELAA is split into $M$ sub-arrays each having $N$ antenna elements. {\color{black} To further enhance the training speed, instead of employing a pure ML-based scheme with no prior knowledge of the channels, as is the case in \cite{monemi20236GFresnel}, as seen in Fig. \ref{fig:ML}, we propose that a rough estimation of a subset of channels denoted by $\widetilde{h}_m$ be leveraged as the input to each SPA $m$.}
It is seen here that for each sub-array $m$, in addition to the measured power $p_m$ and (possibly) estimated CSI vector $\widetilde{\hbold}_m$, the arrival phase $\theta_m$ is also required by the DRL to align all elements $\widetilde{\phi}_{mn}$ through the phase synchronization block. 
This block subtracts the offset $\theta_m$ from all elements of $\widetilde{\phi}_{mn}$ and obtains the new beamforming elements $\phi_{mn}$ so that all radiated beams are added coherently and constructively at the UE location, leading to the desired SBF directivity pattern. 


In practice, all SPAs have identical structures. Noting the domain alignment of the sub-arrays, having single or multiple trained \textit{teacher} SPAs, the remaining SPAs do not require learning from scratch, rather, different schemes of \textit{transfer learning} can be applied to speed up the learning process of remaining \textit{student} SPAs. A similar idea can be employed for learning the whole ELAA for a new DFP when it has been previously trained for some other nearby DFPs. This can effectively handle the mobility management of the UE. 
{\color{black}
\subsection{Mobility and dynamics of the channels}
The interval related to the validity of estimated channels and computed SBF beamformer matrix is limited to the channel coherence time, which depends on the dynamics (i.e., mobility) of the DFP, as well as the propagation environment and operating frequency. For the ideal line-of-sight (LoS) channel model and known DFP location, the CSI can be obtained through simple channel reconstruction methods. Otherwise, especially in highly dynamic channel scenarios, fast CS estimation and beamforming techniques become essential. Two strategies can be considered: (1) applying conventional on-grid/off-grid CS with a reduced number of angular and polar domain samplings; (2) increasing the number of subarray modules and RF-chains with each RF-chain supporting fewer antenna elements, enabling independent channel estimation for each subarray with reduced CS overhead. The latter (former) provides a higher (lower) performance at the cost of higher (lower) hardware and resource costs. 

}
{\color{black}
\section{Conclusion}
In this paper, we studied the technical aspects of electronically controlled near-field spot beamfocusing through ELAAs as well as its applications in various domains of wireless technology. To fully realize the potential of this technology, future research should prioritize the development of efficient SBF algorithms, exploration of novel applications such as meta-lens-based medical solutions, and optimization of system parameters (antenna type and structure, aperture dimensions, operating frequency, etc.) tailored to specific use cases. The scattering and absorption of the EM signals in different tissues over different frequencies can be a challenging issue in realizing SBF for medical and healthcare applications, which needs careful exploration.}


	\bibliographystyle{IEEEtran}
	\bibliography{Mybib}

\end{document}